\documentclass[referee]{raa}            % referee version: for submission

\usepackage{graphicx,times}             %for PS/EPS graphics inclusion, new
\usepackage{amsmath}
\usepackage{times}
\usepackage{graphicx}
\usepackage{graphicx}
\usepackage{epstopdf}
\usepackage{amsmath}
\usepackage{breqn}
\usepackage{color}
\usepackage{soul}
\usepackage{rotating}
\usepackage{amssymb}

\newcommand{\beq}{\begin{equation}}
\newcommand{\eeq}{\end{equation}}
\newcommand{\bea}{\begin{eqnarray}}
\newcommand{\eea}{\end{eqnarray}}
\newcommand{\grb}{GRB~170817A}
\newcommand{\gr}{$\gamma$-ray}
\newcommand{\grs}{$\gamma$-rays}

\def\apj{{ApJ}}
\def\mnras{{MNRAS}}
\def\aa{{A\&A}}

\def\etal{{et al.~}}

\begin{document}

   \title{GRB~170817A: a short GRB seen off-axis
%\,$^*$
%\footnotetext{$*$ Supported by the National Natural Science Foundation of China.}
}
%   \subtitle{I. Place Your Subtitle Here}

   \volnopage{Vol.0 (200x) No.0, 000--000}      %%preserved for Editor. DOn't remove!
   \setcounter{page}{1}          %%starting page, preserved for Editor. DOn't remove!

   \author{Xin-Bo He
      \inst{1}
   \and Pak-Hin Thomas Tam
      \inst{1}
   \and Rong-Feng Shen
      \inst{1}
   }
%% Here is an example of three authors come from different institutes.
%% For single author or all the authors from an institute, use "\inst{}" only

   \institute{School of Physics and Astronomy, Sun Yat-sen University, Zhuhai 519082, China; {\it tanbxuan@mail.sysu.edu.cn}\\
%% Please give the E-mail address of the author, to whom future correspondence and
%% offprint requests will be sent.
          }

   \date{Received~~2009 month day; accepted~~2009~~month day}

\abstract{ 
%It has been commonly believed that the prompt emission of short GRBs comes from a relativistic jet with small half opening angle. However, 
The angular distribution of GRB jets is not yet clear. 
The observed luminosity of GRB 170817A is the lowest among all known short GRBs, which is best explained by the fact that our line of sight is outside of the jet opening angle, $\theta_{obs}$ $>$ $\theta_j$, where $\theta_{obs}$ is the angle between our line of sight and jet axis. Inferred by gravitational wave observations, as well as radio and X-ray afterglow modeling of GRB 170817A, it is likely that $\theta_{obs}$ $\sim$ 20$^{\circ}$ -- 28$^{\circ}$. In this work, we quantitatively consider two scenarios of angular energy distribution of GRB ejecta: top-hat jet, and structured jet with a power law index $s$. For top-hat jet model, we get a large $\theta_j$ (e.g., $\theta_{j}$$>$$10^{\circ}$), a rather high local (i.e., z$<$0.01) short GRB rate $\sim$8--15$\times10^3$ $Gpc^{-3}$$yr^{-1}$ (estimated 90$\sim$1850 $ Gpc^{-3}$$yr^{-1}$ in Fong et al. 2015) and an extremely high $E_{peak,0}$(on-axis,rest-frame)$>$7.5$\times10^4$keV($\sim$500keV for typical short GRB). For structured jet model, we use $\theta_{obs}$ to give limits on $s$ and $\theta_{j}$ for a typical on-axis luminosity short GRB (e.g., $10^{49}$ erg s$^{-1}$$\sim$$10^{51}$ erg s$^{-1}$), and a low on-axis luminosity case (e.g., $10^{49}$ erg s$^{-1}$) gives more reasonable values of $s$. A structured jet model is more feasible for GRB 170817A than a top-hat jet model, due to the rather high local short GRB rate and the extremely high on-axis $E_{peak,0}$ almost rules out the top-hat jet model. GRB 170817A is likely a low on-axis luminosity GRB ($10^{49}$ erg s$^{-1}$) with a structured jet.
% the $\theta_{obs}$ can give a strong limit on $s$ and $\theta_{j}$ for a typical on-axis luminosity short GRB (e.g., $10^{49}$ erg s$^{-1}$$\sim$$10^{51}$ erg s$^{-1}$). 
}

   \authorrunning{X.-B. He, Tam. Thomas \& R. -F. Shen}            %author_head in even pages
   \titlerunning{Off-axis short GRB~170817A }  % title_head in odd pages
    \maketitle
%% The author head (on even pages) and the title head (on odd pages) will be
%% automatically extracted from \author{} and \title{}. Whenever the title is too long,
%% you will be asked to supply a shorter one by inserting either \authorrunning{} or
%% \titlerunning{} before \maketitle. Anyway, you can specify your own heads.
%%
%%
%% Note: In the following text body of your manuscript, please note several differences from
%%       other major journals:
%% (1) \subsection{Please Capitalize the First Letter of Each Notional Word in Subsection Title}
%% (2) Please Capitalize the First Letter of Each Notional Word in all tables' captions

%
%_______________________________________________ sections below
%
\section{Introduction}           %% first-level sections will be auto-capitalized
\label{sect:intro}

Gamma-ray bursts (GRBs) are the brightest flashes of \grs~thought to arise from stellar-level explosions. The duration of observed \gr~emission vary from tens of milliseconds to thousands of seconds. It is well known that the observed duration of GRBs has a bimodal distribution: %The peak is T$_{90}$ $\sim$ 0.2 for short GRBs and T$_{90}$ $\sim$ 20s for long GRBs(Kouveliotou_1993}. 
short GRBs last $\lesssim$ 2s and have harder spectra, while the duration of long GRBs is $\gtrsim$ 2s and their spectra are softer (Nakar, 2007) \footnote{however, there is no clear cut in duration that separate the two classes: there are short GRBs having a duration $>$2s, vice versa.}. Long GRBs are due to collapse of massive stars, and so in most cases they are accompanied by observed supernovae if they are close enough, with some exceptions (e.g., Della Valle \etal 2006). Short GRBs can be produced during mergers of two compact objects, such as two neutron stars or a neutron star with a black hole (Eichler \etal 1989).

Over the last three decades, nearly a thousand short GRBs were discovered by monitoring satellites like BATSE (Fishman \etal 1994), Swift (Gehrels \etal 2004) and Fermi (Meegan \etal 2009). Before GW170817 no gravitational wave (GW) signal was detected from the direction of any short GRBs by LIGO/Virgo (e.g., Abbott \etal 2016a).
%, so compact object mergers as the origin of short GRBs was yet to confirm. 
In recent years, a strong evidence for mergers has been emerging, namely the detection of kilonovae or macronovae in GRB~130603B (e.g., Tanvir \etal 2013, Berget 2014), GRB~060614 (e.g.,Yang \etal 2015, Jin \etal 2015) and GRB~050709 (Jin \etal 2016). Kilonovae or macronovae are powered by r-process nucleosynthesis, a process that can be triggered by NS-NS mergers (e.g. Li \& Paczynski 1998, Kulkarni 2005, Hotokezaka \etal 2013). With the Fermi GBM detection of GRB 170817A after the aLIGO event GW170817 (LIGO Scientific Collaboration \& Virgo Collaboration 2017a), the first direct evidence of a NS-NS merger origin for short GRBs has been established.

In the dawn of GW astronomy, short GRBs are one of the best electromagnetic wave counterparts of GW events (e.g. Baiotti \& Rezzolla 2017, Paschalidis 2017). Since Fermi was launched in 2008, the GBM has detected more than 350 short-duration GRBs (Gruber \etal 2014). The GBM consists of 12 NaI detectors with energy range of 8~keV to 1~MeV, and 2 BGO detectors with energy range of 200~keV to 40~MeV. When Fermi GBM triggered a GRB, the GRB location can be calculated from the 12 NaI detectors. In the current multi-messenger astronomy era, it is foreseen that more and more low luminosity GRBs like GRB 170817A will be seen by GRB monitors in the future including SVOM, GeCAM, etc.; many of these GRBs may be EM counterparts of GW events, and those GRBs may help us to obtain accurate GW source locations, paving the way for studying merger environment (e.g., host galaxy types) and their progenitors. %The observation of GW170817 show that it is consistent with GRB 170817A being produced by the merger of NS-NS in NGC 4993, it's a exciting observation, then we have a multi-messenger observation from radio to $\gamma$-ray of this event was reported, e.g.(Abbott \etal 2017). 
In this paper, we analyze the Fermi GBM's data of GRB 170817A and compare its properties with the short GRB population. We argue that an off-axis line of sight in different jet models can quantitatively explain its low luminosity in $\gamma$-ray band.
%\add{we confirm it is a typical short GRB. And according to the lowest observed E$ _{iso,\gamma}$, we argue three different jet model of the off-axis emission to explain its low observed $\gamma$-ray luminosity}.
%% Authors can give a citation as 'Michel et al. 1992'.
%% You may also use \cite, \citep and \citet for citation, and use Table~1 or Figure~1
%% and so forth. Using \ref and \label for cross-references of Tables/Figures
%% is a good way in adjusting/adding/removing text, tables or figures.
\section{Properties and data analysis of \grb}
\label{sect:Properties}
\subsection {Properties of \grb}
GRB 170817A triggered Fermi-GBM instruments at 12:41:06.475 UT on 17 August 2017 (which is defined as $t _{0}$ in this work; von Kienlin \etal 2017, see Figure~\ref{light_curve_GBM}) with a time delay of $\sim$ 1.7s after GW170817 triggered. The INTEGRAL/SPI-ACS also detected this short GRB (Savchenko \etal 2016), removing any doubt on the reliability of the GBM detection. The time duration of GRB 170817A, is T$_{90}$=1.984$\pm$0.466~s in the 50--300 keV band\footnote{from the Fermi GBM Burst Catalog, online version available at https://heasarc.gsfc.nasa.gov/W3Browse/fermi/fermigbrst.html, Gruber, D. et al. 2014, von Kienlin, A. et al. 2014, Bhat, P. et al. 2016}.
The GBM-determined location is RA = 176.8, DEC = -39.8 (J2000 degree) by on-ground calculation, with an uncertainty of 11.6 degree (1-sigma containment; von Kienlin \etal 2017). Figure~\ref{light_curve_GBM} shows the count rate seen by the Fermi GBM detector. It can be seen that the light curve shows a weak and short pulse. The Fermi LAT did not detect this GRB, the angle from Fermi LAT boresight is 91 degree at the GBM trigger time (von Kienlin \etal 2017)
%It can be seen that the NaI 2 light curve shows a weak and short pulse, mainly at energies below 100~keV, while the BGO light curve does not show any enhancement above the background level. The Fermi LAT did not detect this GRB, the angle from Fermi LAT boresight is 91 degree at the GBM trigger time (von Kienlin \etal 2017)

\subsection {Fermi GBM spectral analysis of \grb}

 \begin{figure}
\includegraphics[width=7.5cm, angle=0]{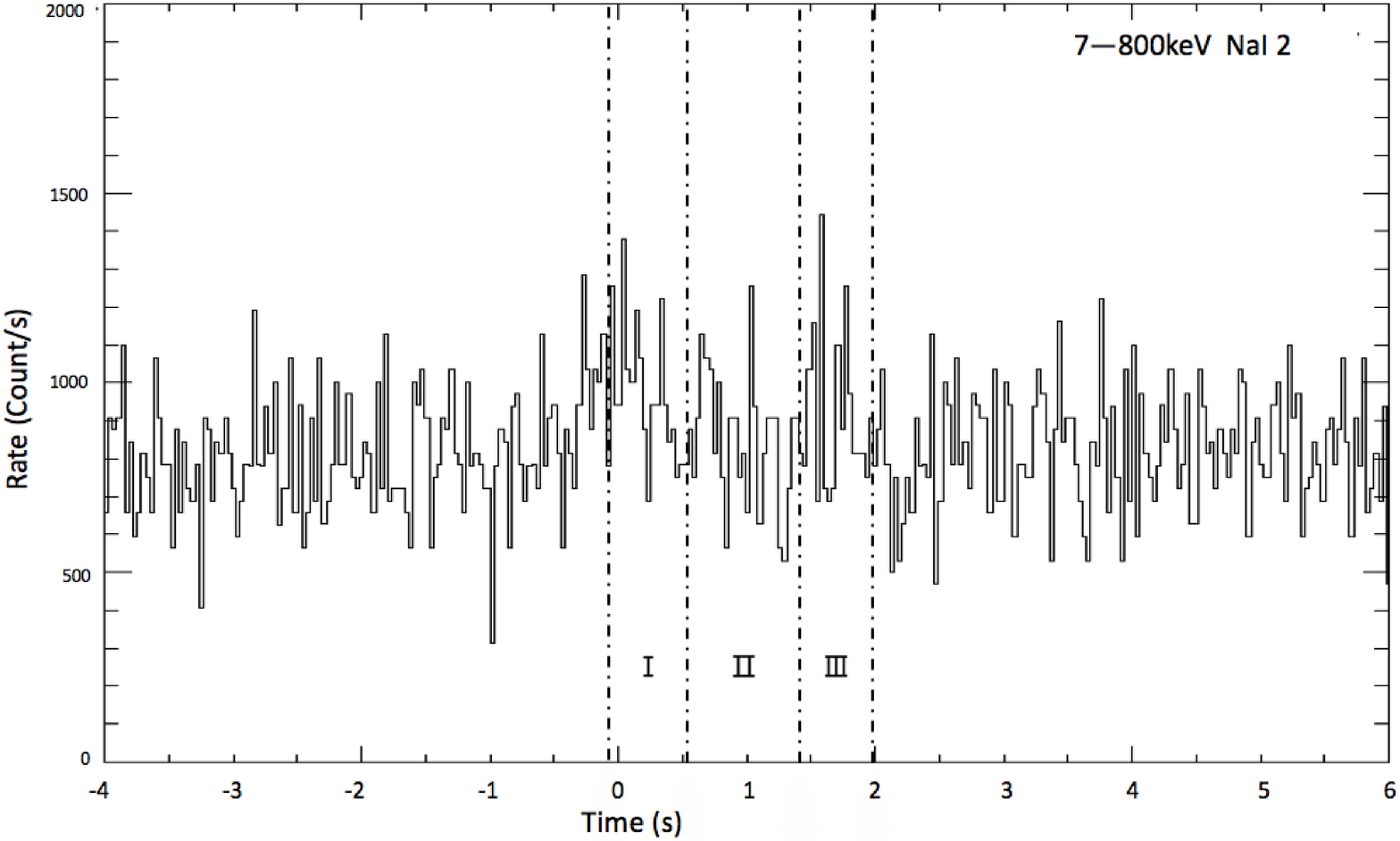}
\includegraphics[width=7.5cm, angle=0]{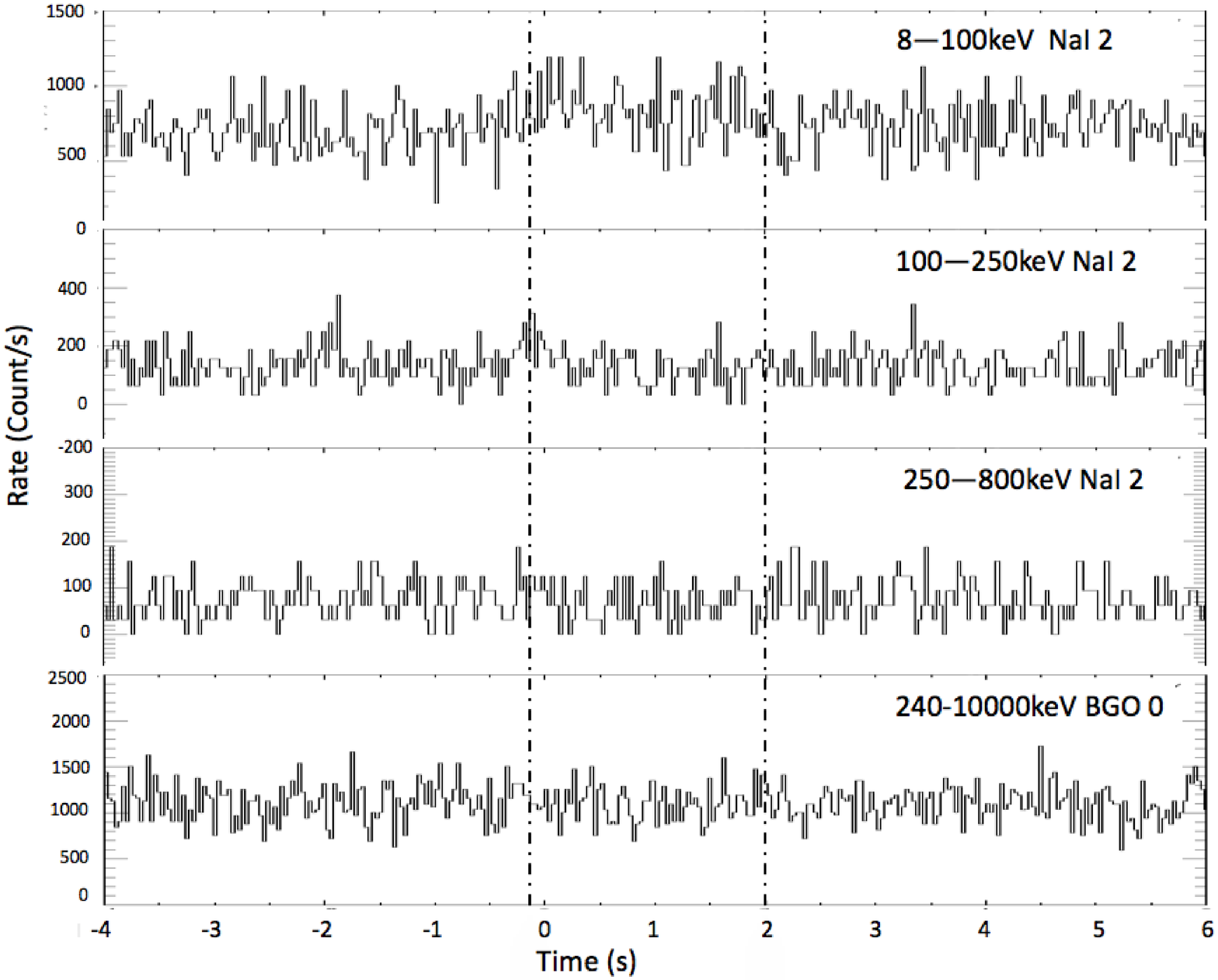}
\caption{\emph{Left panel:} The 7--800~keV light curve of \grb~obtained by the Fermi NaI 2 detector with 32ms temporal resolution. Periods I, II, and III indicate different time ranges for spectral analysis (Sect. 2.2). \emph{Right panel:} Light curves of \grb~observed by Fermi NaI 2 detector binned with the same temporal resolution in different energy ranges. The two dashed-dotted lines mark the period used in the whole burst spectral analysis as shown in Table 1.}
 \label{light_curve_GBM}
\end{figure}

We selected the data obtained by NaI(1,2,5) detectors whose pointing direction is within a burst angle of 60 degrees, available at the Fermi Science Support Center. We analyzed the time tagged event (TTE) data, taking advantage of its 32ms timing resolution, with the RMFIT 43pr2 package. For spectral analysis, we use three common spectral models: power-law (PL), Comptonized model (COMP), and Black-Body(BB), to fit the data. The PL model is defined as:
\begin{equation}
\label{eqn:PL}
f_{PL}(E) = A (\frac{E}{E_{piv}})^{-\alpha},
\end{equation}
where A is the normalization factor at 100 keV in unit of ph~s$^{-1}$~cm$^{-2}$~keV$^{-1}$, $\alpha$ is the spectral index and $E_{piv}$ is fixed to 100 keV. 

Some short GRBs have a thermal component from the photosphere emission which can produce black body radiation, and thermal radiation may also be expected for magnetar bursts (soft gamma-ray repeaters), so we also fit the spectra with the model of a Black-Body:
%(Pe'er \etal 2006, Giannios 2008, Beloborodov 2010).
\begin{equation}
\label{eqn:BB}
f_{BB}(E) =A\frac{E^2dE}{(kT)^{4}exp(E/kT)-1},
\end{equation}
where A and kT are free parameters. Here A is the normalization factor in unit of erg~cm$^{-2}$~s$^{-1}$ and kT is in unit of keV.
The COMP model is a useful model for short GRBs, which is represented by a power law with an exponential cutoff:
\begin{equation}
\label{eqn:COMP}
f _{COMP}(E) = A(\frac{E}{E_{piv}})^\alpha \exp\left[-\frac{(\alpha+2)E}{E_{peak}}\right].
\end{equation}
The free parameters are A which is amplitude factor at 100 keV in unit of ph~s$^{-1}$~cm$^{-2}$~keV$^{-1}$, $\alpha$ is the low-energy spectral index, and E$ _{peak}$ is the peak energy in units of keV. $E_{piv}$ is also fixed to 100 keV. 
%\begin{figure}
%\centering
   % \includegraphics[width=8.5cm, angle=0]{fg1.eps}
     % \caption{the light curve of GRB~170817A seen by the INTEGRAL SPI-ACS detector (retrieved from http://www.isdc.unige.ch).}
        % \label{light_curve_INTEGRAL}
  % \end{figure}

Using RMFIT for spectral analysis, we consider four time intervals (see Table~1 and Figure~\ref{light_curve_GBM}). For every interval, we try the above three models to fit the data. The PL model can satisfactorily fit the spectrum in all intervals. In the first interval $t _{0}-$0.128s to $t _{0}+$0.512s and the main burst interval $t _{0}-$0.128s to $t _{0}+$1.984s, the spectrum is better fit by the COMP model with improvement of $\Delta$ C-STAT $>$ 5. The results of the spectral analysis are summarized in Table 1. Our analytic result is consistent with Fermi GBM team's result (Goldstein \etal 2017). 

\begin{figure*}
\includegraphics[width=8.5cm, angle=0]{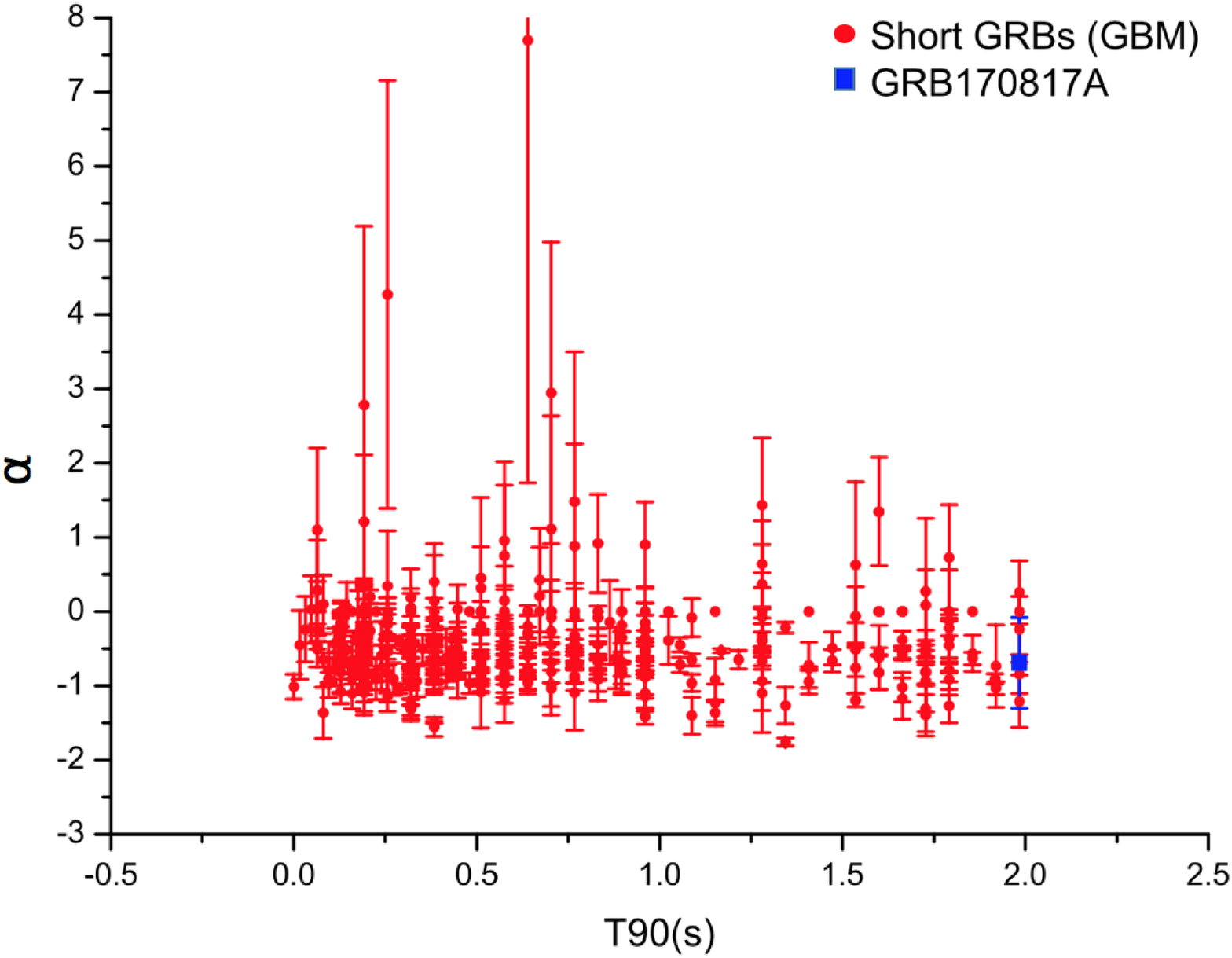}
\includegraphics[width=8.5cm, angle=0]{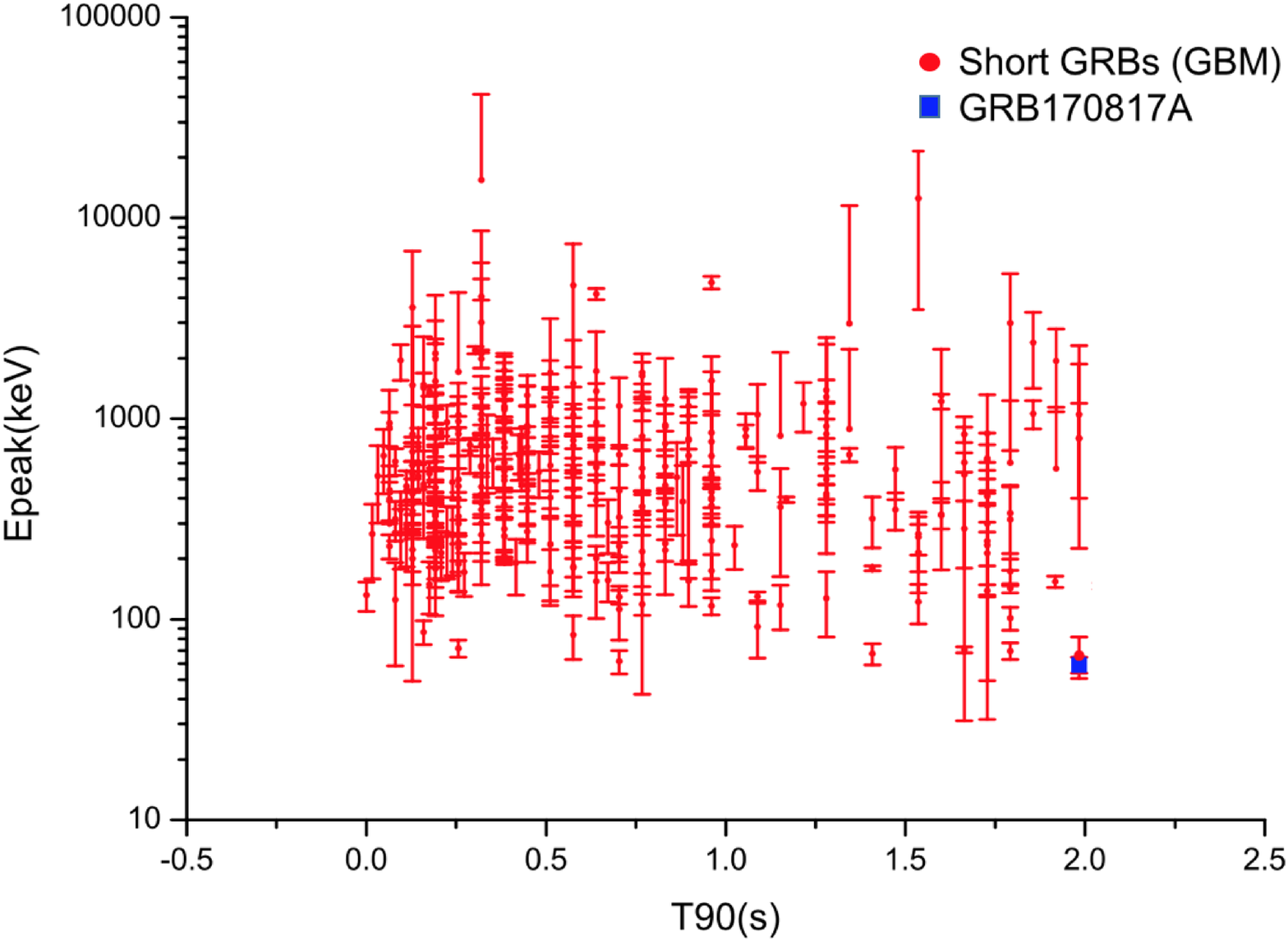}
\includegraphics[width=8.5cm, angle=0]{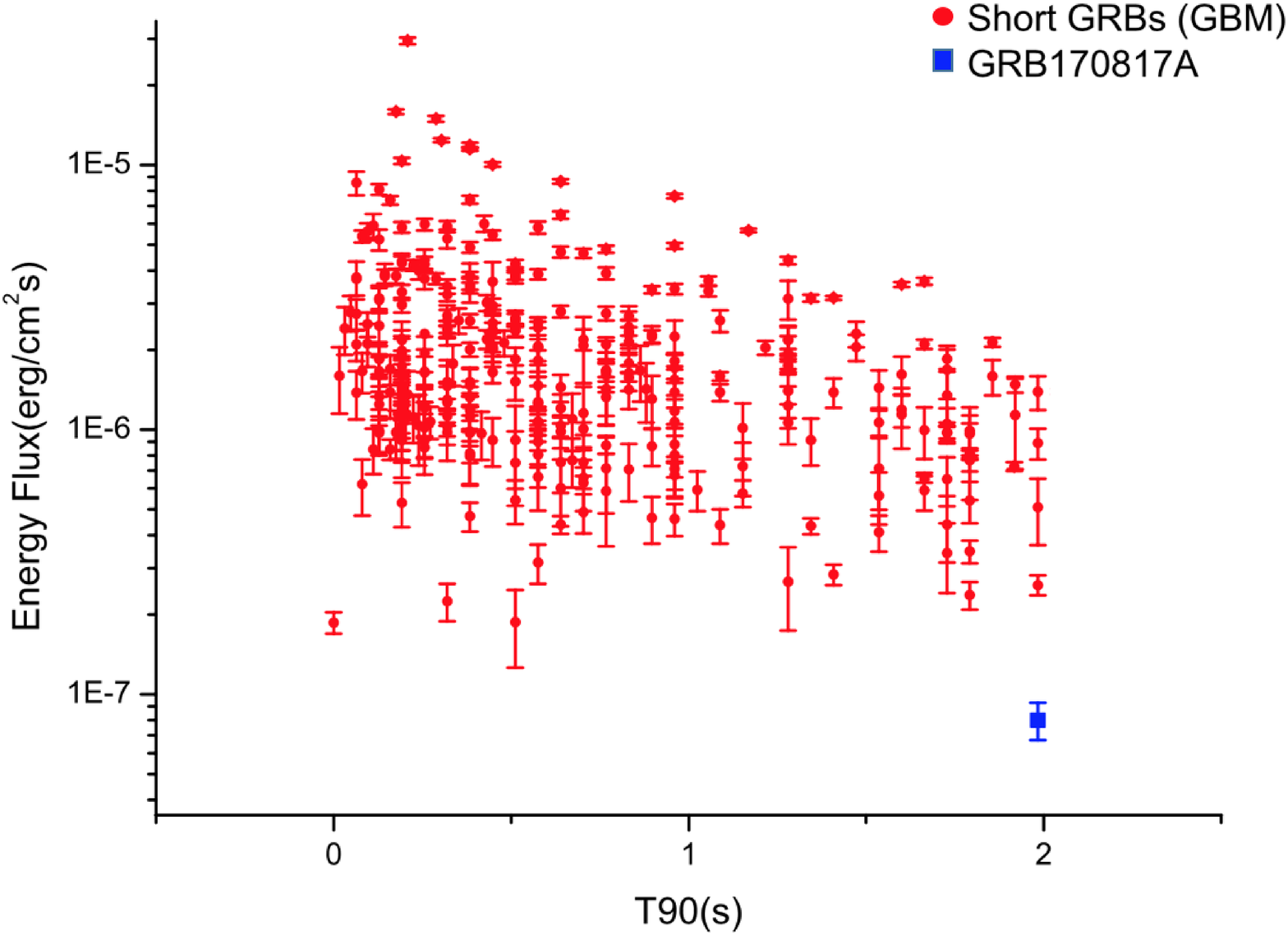}
\includegraphics[width=8.5cm, angle=0]{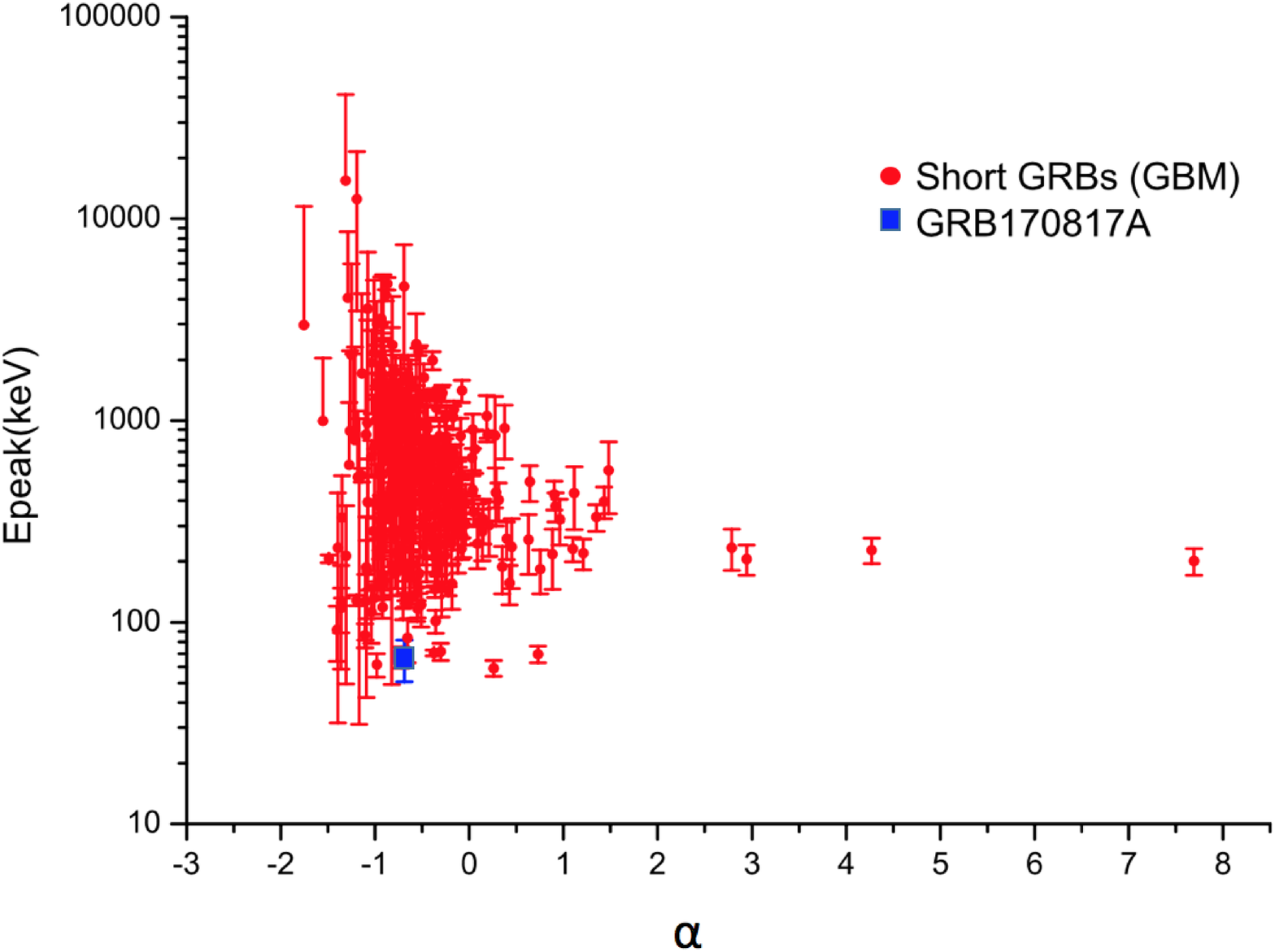}
 \caption{ The upper-left, upper-right, and lower-left figures show T90 plotted against three spectral parameters obtained from the COMP model fits: low-energy index $\alpha$, $E_{peak}$, and average energy flux, respectively. Red points represent GBM-detected short GRBs (T90$<$2.0s, since 2008) which can be satisfactorily fitted by the COMP model over the duration of the burst. The blue point represents GRB 170817A.}
         \label{T90-parameter}
       %\caption{\emph{eft:)}TTE light curve of GRB 170817A, the time bins is 32ms in NaI 2 detector. The dotted lines are the background region for fitting, the hatching is the source region for T90 analysis. {\emph{right:)} The energy fluence evolution of NaI 2 detector. The vertical dashed lines are the times the energy flucene is 5\%, 25\%, 75\%, 95\% of total energy fluence. The time range defining the T$ _{50}$ and T$ _{90}$. }
   \end{figure*}

Now we are ready to compare the properties of GRB~170817A with those of other short GRBs. The low-energy index, $\alpha$, of GRB 170817A in the COMP model is similar to other short GRBs, c.f., the average $\alpha _{SGRB}$ = $-$0.6 ($\sigma$ = 0.4; Avanzo \etal 2014). %We also calculated the spectral hardness of GRB 170817A. Following the steps in Narayana \etal (2016), we summed the background-subtracted photon counts in each detector during T$_{90}$, and we selected the two energy ranges: 10-50 keV and 50-300 keV (von Kienlin, A. \etal 2014). In our result, the hardness ratio is within the range of most short GRBs as shown in Fig. 6 of von Kienlin \etal (2014), GRB 170817A has a formally hard spectrum (Kouveliotou \etal 1993). Thus we regard GRB 170817A as being a short-hard GRB. %The $E _{peak}$ as shown in Table 1 are below 200 keV.  The main emission interval is -0.112 to 0.504 which can be fit with a Power Law model. (Xinbo: please rewrite this following sentence) %%%With the hardness ratio of $\approx$ 1.2, which is within the range of 1 to 10 for most short GRBs With the hardness ratio of $\approx$ 1.2, which is within the range of 1 to 10 for most short GRBs 
We compare the spectral parameters of GRB 170817A with those of GBM-detected short GRBs satisfactorily fitted by the COMP model over the $T_{90}$ duration. To do this, we obtained the relevant COMP spectral parameters of these short GRBs (taken as  $T_{90}\leq2$~s) from the GBM spectral catalog (Gruber, D. \etal 2014, von Kienlin, A. \etal 2014, Bhat, P. \etal 2016). The results are shown in Figure~\ref{T90-parameter}. It can be seen that GRB 170817A still lies within the main distribution in the two figures plotting T90 versus $\alpha$ or $E_{peak}$. In contrast, in the plot showing the average energy flux versus T90, GRB~170817A is clearly an outlier. Its average energy flux in the COMP model is 0.8$\times$10$^{-8}$erg/(s$\cdot$cm$^2$), lower than Fermi GBM short GRBs. And its energy fluence is $\sim$2$\times$10$^{-7}$erg/cm$^2$. From the above comparison, GRB 170817A is a low-luminosity short GRB. 
%\begin{figure}
%\includegraphics[width=8.5cm, angle=0]{T90-32ms-source.eps}
%\includegraphics[width=8.5cm, angle=0]{T90-32ms-fluence.eps}
%\caption{\emph{(Top:)} TTE light curve of GRB 170817A, the time bins is 32ms in NaI 2 detector. The dotted lines are the background region for fitting, the hatching is the source region for T90 analysis. \emph{(Bottom:)} The energy fluence evolution of NaI 2 detector. The vertical dashed lines are the times the energy flucene is 0\%, 5\%, 25\%, 75\%, 95\%, and 100\% of the total energy fluence. The time range defining the T$ _{50}$ and T$ _{90}$. }
% \label{light_curve_GBM}
%       %\caption{\emph{eft:)}TTE light curve of GRB 170817A, the time bins is 32ms in NaI 2 detector. The dotted lines are the background region for fitting, the hatching is the source region for T90 analysis. {\emph{right:)} The energy fluence evolution of NaI 2 detector. The vertical dashed lines are the times the energy flucene is 5\%, 25\%, 75\%, 95\% of total energy fluence. The time range defining the T$ _{50}$ and T$ _{90}$. }
%  \label{T90}
%   \end{figure}
\section{Discussion}
It has been reported in the literature that GRB~170817A locates in the nearby galaxy NGC~4993, e.g., Coulter \etal (2017). The redshift of NGC~4993, and therefore GRB 170817A, is z $\approx$ 0.0098, which corresponds to a luminosity distance of 42.5 Mpc (assuming $H_0$ = 69.6, $\Omega_{M}$=0.286, $\Omega_{vac}$=0.714). Therefore, GRB 170817A is the nearest short GRB with known redshift (it is also the second nearest GRB, after GRB~980425 that lies at z=0.0085; Galama \etal 1998). 
\begin{table*}
\begin{minipage}[]{200mm}
\caption{Model fits of the main emission episodes. \label{gbm_spec}}
%\begin{rotatetable}
\begin{tabular}{l@{ }c@{  }c@{  }c@{  }cccc@{ }c}%??????
    \hline
    \hline
    $t-T_{0}$~(sec) & $Model$ & $E_{peak} (keV)$ & $\alpha$ & $kT(keV)$ & $c-stat/dof$ & $photon  flux^{a}$ & $energy  flux^{b}$  & $\Delta c-stat$\\
    \hline\hline\noalign{\smallskip}
-0.128--0.512  & PL & -- & -1.59$\pm$0.119 & -- & 460.3/407 & 2.16$\pm$0.33 & 2.99$\pm$0.55& -- \\\hline
  & BB & -- & -- & 30.73$\pm$4.48 & 468.9/407 & 1.12$\pm$0.18 & 1.52$\pm$0.26& -8.6 \\\hline
  & COMP & 194.3$\pm$ 112.0 & -1.01$\pm$0.41 & -- & 455.8/406 & 2.00$\pm$0.33 & 2.40$\pm$0.71& 4.5 \\\hline
0.512--1.408  & PL & -- & -1.98$\pm$0.46 & -- & 529.7/407 & 0.68$\pm$0.26 & 0.52$\pm$0.36& -- \\\hline
  & BB & -- & -- & 11.88$\pm$2.65 & 524.1/407 & 0.73$\pm$0.22 & 0.41$\pm$0.13& 5.6 \\\hline
  1.408--1.984  & PL & -- & -2.19$\pm$0.346 & -- & 447.6/407 & 1.45$\pm$0.34 & 0.86$\pm$0.39& -- \\\hline
  & BB & -- & -- & 10.22$\pm$1.77 & 445.9/407 & 1.27$\pm$0.30 & 0.63$\pm$0.16& 1.7 \\\hline
\hline 
 -0.128--1.984  & PL & -- & -1.81$\pm$0.133 & -- & 487.5/407 & 1.32$\pm$0.18 & 1.29$\pm$0.27& -- \\\hline
  & BB & -- & -- & 12.70$\pm$1.23 & 485.5/407 & 1.08$\pm$0.15 & 0.65$\pm$0.08& 2.00 \\\hline
  & COMP & 66.06$\pm$ 15.5 & -0.69$\pm$0.61 & -- & 481.6/406 & 1.24$\pm$0.17 & 0.80$\pm$0.13& 5.9 \\\hline
\end{tabular}
%\end{rotatetable}

a:10--1000~keV,in units of photons/(s$\cdot$cm$^2$);  

b:10--1000~keV,in units of $\times$10$^{-7}$erg/(s$\cdot$cm$^2$); 

PL is PowerLaw model, BB is Black-Body model, COMP is Comptonized,Epeak model
\end{minipage}
\end{table*}

As inferred from its observed energy fluence of $\approx$2$\times$10$^{-7}$~erg~cm$^{-2}$, the isotropic-equivalent energy in \grs, $E _{iso,\gamma}$, can be calculated by:
\begin{equation}
\label{eqn:Energy}
E _{iso,\gamma} = \left(\frac{4{\pi}D^2_{L}F }{1+z} \right),
\end{equation}
where $D _{L}$ is the luminosity distance and F is the total energy fluence of GRB~170817A. Therefore, E$ _{iso,\gamma}$ $\approx$ 4 $\times$ 10$^{46}$ erg, which is significantly lower than those of typical short GRBs (i.e., E$ _{iso,\gamma}$~$\gtrsim$10$^{50}$~erg). Correspondingly, GRB~170817A has a low luminosity $L_{iso,\gamma} \approx  2\times10^{46}$ erg s$^{-1}$. Such a low luminosity is the most striking feature of GRB 170817A. For comparison, previously detected bright short GRBs typically have $L_{iso,\gamma} \approx 10^{51}$ erg s$^{-1}$ (Zhang \etal 2012).

GRBs come from the relativistic jet, but the angular energy distribution within the jet (or ejecta) is not yet known. It is well known that $L_{obs}$ strongly depends on the viewing angle, $\theta_{obs}$, and for a GRB with a given intrinsic on-axis luminosity, $L_{obs}$ decreases with $\theta_{obs}$.
%If the line of sight is in the jet, the $L_{obs}$ is very low for a short GRB, so we can rule out GRB 170817A is an on-axis short GRB. We can conclude that the line of sight is out of the jet, so the $L_{obs}$ is much lower than typical short GRBs, in this case, the GRB 170817A is an off-axis short GRB. The $L_{obs}$ is strongly related with $\theta_{obs}$. \del{Its}\add{It is} lucky for us that we have multi-band observation of this event, it can help us to limit the value of $\theta_{obs}$ by other band observation. 
From the detected GW170817, $\theta_{obs}$ $<$ $28^{\circ}$ is implied by LIGO's data (Abbott et al. 2017). Margutti et al. (2017) infer $\theta_{obs}$ $\sim$ $20^{\circ}$ -- $40^{\circ}$ from radio and X-ray afterglow observations. Combining the results from these two papers, we focus on the discussion of off-axis gamma-ray emission to be $\theta_{obs}$ $\sim$ $20^{\circ}$ -- $28^{\circ}$.
Next we consider two widely discussed models describing the angular distribution of energy within the GRB ejecta: top-hat jet and structured jet.%We applied it to  of GRB 170817A as an off-axis short GRB,  respectively. All of those can explain the low luminosity of GRB 170817A which the line of sight is out of the jet.
%There are two major models to describe the energy distribution of the GRB ejecta. 
%The first one assumes uniform energy distribution which the short GRBs called off-beam GRBs. The other one is the non-uniform energy distribution which the short GRBs called off-axis GRBs. 

\subsection {Top-hat jet model for GRB~170817A}

In the standard fireball model, the angular energy distribution is uniform in the jet, and outside the jet, the energy is close to zero (Piran 1999). In this case, to be able to see the GRB, the line of sight must lie within the jet half-opening angle $\theta_j$, or with slight offset with the typical Lorentz factor $\sim$ 200 at the prompt phase (Salafia \etal 2015). For a GRB with a typical on-axis luminosity, say, $10^{51}$ erg s$^{-1}$, and the observed luminosity (to our line of sight) is as low as $L_{obs} \approx 10^{46}$ erg s$^{-1}$, $\theta_{obs}$ should be about $\sim$ 2$\theta_j$(see Figure 3, Salafia \etal 2015) with a typical Lorentz factor of 200. Applying this to the case of GRB~170817A, the observed angle should be 
%$\theta_j$ $\lesssim$ $\theta_{obs}$ $\lesssim$ 2$\theta_j$. 
%But GRB~170817A is typical short GRB, the jet angle should not be very large, so in our analysis, we think that the 
$\theta_{obs}$ $\approx$ 2$\theta_j$.
 Considering that $\theta_{obs}$ $\sim$ $20^{\circ}$ -- $28^{\circ}$, we have $\theta_{j}$ $\sim$ $10^{\circ}$ -- $14^{\circ}$ with a typical Lorentz factor. This means that the jet half-opening angle $\theta_{j}$ is consistent with previously inferred for short GRBs (e.g., $\sim$$16^{\circ}$; Fong et al., 2015; though there are lower estimates, see Jin et al. 2017). 

Taking the above values, the local (z$\leq$0.01) rate of short GRBs can be estimated by:
\begin{dmath}
\label{eqn:rate1}
\qquad R _{nus} = \left(\frac{N _{event} }{V(z \leq 0.01)T } \right)\left(\frac{4\pi}{FoV} \right)\left(\frac{1}{1-cos(2\theta_j)}\right)  
%\approx 326_{-147}^{+216} \times10^{3}Gpc^{-3} yr^{-1}\left(\frac{\theta_j}{0.1rad}\right)^{-2} N _{event}\left(\frac{T}{9yr}\right)^{-1}\left(\frac{9.5sr}{F.o.V} \right),
\end{dmath}
where $N _{event}$ is the total number of short GRBs detected within the comoving volume $V$(z$\leq$0.01) and the observation time span $T$. The Fermi GBM was launched in 2008, thus we take $T= 4.5$ yr (taking fractional effective exposure to be 0.5), and it has a field of view FoV $\approx9.5$ sr. Currently, GRB 170817A is the only detected GBM burst with known $z\leq 0.01$, and we have $N_{event}=1$. Therefore $R _{nus}$ is $\approx 1.5\times10^4$ $Gpc^{-3}$$yr^{-1}$ for $\theta_{obs}$ $\sim$ $20^{\circ}$, $\theta_{j}$ $\sim$ $10^{\circ}$ and $\approx 8.1\times10^3$ $Gpc^{-3}$$yr^{-1}$ for $\theta_{obs}$ $\sim$ $28^{\circ}$,$\theta_{j}$ $\sim$ $14^{\circ}$ in local universe (z$\leq$0.01). This is at the high end of the estimate given in Fong et al. (2015; 90$\sim$1850 $ Gpc^{-3}$$yr^{-1}$). Such a high event rate of local short GRBs may be problematic for the top-hat jet model. 

In the off-axis jet, the peak energy $E_{peak}$ of the observed $vF_{v}$ spectrum varies with the observing angle $\theta_{obs}$. We can calculate the (on-axis, rest-frame) peak spectral energy $E_{peak,0}$ by the observed $E_{peak}$. The peak energy $E_{peak,0}$ can be estimated by (Salafia et al. 2016):
\begin{equation}
E_{\rm peak}(\theta_{obs})=\frac{E_{\rm peak,0}}{1+z}\times\left\lbrace\begin{array}{lr}
                                                    1 & \theta_{obs} \leq \theta_j\\
                                                    \frac{\delta_B}{(1+\beta)\Gamma} & \theta_{obs} > \theta_j\\
                                                   \end{array}\right.
\end{equation} 
From the Fermi data analysis in Sec 2.2, the observed $E_{peak}$ of the $T_{90}$ duration is $\approx$ 66 keV in Table 1, the z $\approx$ 0.0098, the doppler factor is defined as $\delta_{\rm B} = \Gamma^{-1}\left[1-\beta\cos\left(\theta_{obs}-\theta_j\right)\right]^{-1}$, a typical $\Gamma$ $\approx$ 200, so the on-axis peak energy $E_{peak,0}$ should be $>$$7.5\times10^4$ keV for $\theta_{obs}$ $\sim$ $20^{\circ}$--$28^{\circ}$, $\theta_{j}$ $\sim$ $10^{\circ}$--$14^{\circ}$ in rest-frame. This $E_{peak,0}$ is too large for a typical short GRB($\sim500$keV) in Fermi GBM Burst Catalog. So the estimated $E_{peak,0}$ almost rules out the top-hat jet model of GRB 170817A.
%, although larger $\theta_{j}$$>>$$\add{10}^{\circ}$ may circumvent this issue. %In this model, the lower limit of the rate of neutron star mergers (z$\leq$0.01) is $\sim$ 470~Gpc$^{-3}$$yr^{-1}$ ($\theta_{obs}$=$90^{\circ}$). %The above argument is similar to saying that the chance to detect an off-axis short GRB in such a small comoving volume (z$\leq$0.01) is very low even for a 9.0-year mission. 
%So if GRB 170817A has a top-hat model, $\theta_{j}$ may be significantly larger than previously thought (i.e. $\theta_{j}>10^{\circ}$). 

\subsection {Structured jet model for GRB 170817A}
It has been proposed that a structured jet may explain some long GRBs with low luminosity. The structured jet is widely discussed in those scenarios: a power-law distribution model (e.g. Rossi, Lazzati, \& Rees, 2002, Dai \& Gou 2001), a Gaussian-type jet model (e.g. Zhang et al. 2004) and a two component jet model (e.g.  Huang et al. 2004). This idea was also suggested for short GRBs (e.g. Aloy, Janka, \& M\"uller, 2005, Murguia-Berthier \etal 2017). It may therefore be conjectured that GRB 170817A is a typical short GRB with isotropic-equivalent luminosity $L_{core}\sim10^{49}$ erg s$^{-1}-10^{51}$ erg s$^{-1}$ inside the jet core (i.e., within a half opening angle $\theta_{j}$), but our line of sight is out of $\theta_j$. In the structured jet model, the energy does not sharply decrease to zero outside the jet core, but can instead follow a power-law decrease. In this case, the jet luminosity per solid angle along the direction of $\theta_{obs}$ is described by $L _{obs}(\theta_{obs})$=$L _{core}$($\theta_{obs}$ / $\theta_{j})^{-s}$ for $\theta_{obs}>\theta_{j}$ (Pescalli \etal, 2015), such that the angular dependence of the energy distribution outside $\theta_{j}$ is described by the power index, $s$. Different values of $s$ have been acquired through either simulations or observations, and can range from 2 to 8 (e.g., Frail \etal 2001, Pescalli \etal 2015, Kathirgamaraju \etal 2017). 

Here, we assume that the core luminosity in the jet $L_{core}$ is $10^{49}$ erg s$^{-1}-10^{51}$ erg s$^{-1}$. Figure~3 shows the schematic relation between $s$ and $\theta_{obs}$ for GRB~170817A, taking representative values of $\theta_{j}$: 1$^\circ$, 3$^\circ$, 6$^\circ$, 10$^\circ$, and 15$^\circ$. This figure can apply to future short GRBs with similar observed $\gamma$-ray luminosity with different $\theta_{obs}$. 
%Therefore, it is possible to infer possible values of $s$ and $\theta_{j}$ with the estimate of $\theta_{obs}$ of GRB~170817A.
For GRB~170817A, we have $\theta_{obs}$ $\sim$ $20^{\circ}$ -- $28^{\circ}$. In the following, we discuss the correlation between $s$ and $\theta_{j}$ for two typical values of on-axis $L_{iso,\gamma}$:

1) If GRB 170817A is a typical bright short GRB with on-axis $L_{iso,\gamma} \sim 10^{51}$ erg s$^{-1}$, the half-opening angle $\theta _{j}$ is constrained to be $<$10$^{\circ}$ for s $<$ 10, and we can rule out $s<$2 for all $\theta_{j}>$1$^\circ$. If GRB 170817A has $\theta_{j}$ $=$ $6^{\circ}$ (c.f. Jin \etal, 2017), then $s$ should be larger than 7, which is very large. Therefore, an on-axis luminosity of $L_{iso,\gamma} \sim 10^{51}$ erg s$^{-1}$ is not preferred for GRB 170817A.

2) If GRB 170817A is a short GRB with on-axis $L_{iso,\gamma} \sim 10^{49}$ erg s$^{-1}$, then $\theta_{obs}$ $\sim$ $20^{\circ}$ -- $28^{\circ}$ can be satisfied with $\theta_{j}$ up to 15$^\circ$ for s$<$10. On the other hand, the constraints for $s$ are not severe (3$<$s$<$9) for $\theta_{j}=3^{\circ}-10^{\circ}$. Therefore, we conclude that $\theta_{j}$ should be small ($\lesssim$ $15^{\circ}$) for s$<$10, which is not a severe constraint. 

Therefore, a low on-axis gamma-ray luminosity ($L_{iso,\gamma} \sim 10^{49}$ erg s$^{-1}$) is preferred for GRB~170817A in the context of the structured jet model.

\subsection {Comparison with other low-luminosity GRBs}
Another related phenomenon is low-luminosity GRBs (llGRBs; e.g., Liang \etal 2007, Nakar 2015) which includes GRB 060218 and GRB 980425. They have gamma-ray luminosity smaller than typical long GRBs (i.e., $<$10$^{48}$ erg/s). However, their gamma-ray emission properties are different from GRB 170817A. Even compared to conventional long GRBs, llGRBs are longer ($\sim$1000s) and softer ($E _{peak}<$100 keV). The real nature of llGRBs are still unclear, but they are thought to arise from the same progenitors of long GRBs, and are associated with broad-line Type Ic SNe. Only a handful of llGRBs are known, but from the observed rate, they out-number long GRBs in the local Universe by about an order of magnitude. There are also evidence that the beaming factor of llGRBs is larger than long GRBs, so llGRBs are less collimated. A similar situation may apply to GRB 170817A: although this is the first known low-luminosity short GRB (regardless of whether they are intrinsically less luminous or seen off-axis), the true rate might be much higher than short GRBs having more typical luminosity (e.g., L$>$10$^{50}$ erg/s).

\subsection {Low luminosity events -- Burst of a soft gamma-ray repeater}

Another proposed low luminosity events are soft gamma-ray repeaters (SGRs) in the local Universe. These extragalactic giant SGR flares from young magnetars with a long recurrence timescale may mimic a small portion of short GRBs, as was discussed in several cases (Abbott \etal 2008, Ofek \etal 2008, Hurley \etal 2010, Abadie \etal 2012). The peak luminosity of SGR giant flares ranges from $10^{44}$ to $10^{47}$ erg/s. Considering the energetics of GRB 170817A, its low luminosity and E$ _{iso,\gamma}$ are consistent with such events. However, a major uncertainty is that whether a magnetar exists after the merger of the two neutron stars.%Maybe the physical processes of the GRB 170817A are like the magnetar giant flare, they have a same origin. 
The duration of GRB 170817A ($\sim$2~s) is a bit longer than previously seen SGR flares, but the current sample of SGR giant flare light curves is still too small to exclude such a possibility.
%There are different ways to distinguish the magnetar scenario and the NS-NS merger scenario for short GRBs in a nearby Galaxy (e.g., identification of X-ray pulsars, energetics, spectrum), but detectable gravitational wave is expected from NS-NS mergers but not from young magnetars.

\section{Summary}
GRB 170817A is the closest short GRB ever known. The isotropic-equivalent energetics, E$ _{iso,\gamma}$ of GRB 170817A is very low (i.e., its E$ _{iso,\gamma}$ is only 4$\times$10$^{46}$ erg), which is 3-4 orders of magnitude lower than other short GRBs. In our paper, we analyze the GBM data of GRB 170817A, and we confirm that GRB 170817A is a typical short GRB but with low observed gamma-ray luminosity, consistent with most papers in the literature. 
%\del{Because of its low observed luminosity, we consider the case that our line of sight is outside of the jet opening angle, $\theta_{obs}$ $>$ $\theta_j$. The observed luminosity of a GRB depends on the viewing angle, and this may accommodate the observed very low luminosity of GRB~170817A. }
%In our paper, we get that GRB 170817A is more likely a weak short GRB (e.g. $L_{iso,\gamma} \sim 10^{49}$) for the observation data.

Inferred from gravitational wave data and radio-to-X-ray afterglow modeling, we take $\theta_{obs}$ to be $\sim$ $20^{\circ}$ -- $28^{\circ}$. We then compare the top-hat jet model and the structured jet model. According to our analysis, we find that a structured jet model (e.g., Aloyetal \etal 2005, Murguia-Berthier \etal 2017, Kathirgamaraju \etal 2017) is a more feasible model for GRB 170817A than a top-hat jet model. For structured jet model, $\theta_{obs}$ can give a strong limit on s and $\theta_{j}$ for a typical GRB. If GRB170817A is a weak source, the structured jet model can fit the observation (the $\theta_{obs}$ and the low observed luminosity) with reasonable $\theta_j$ and index $s$ (e.g. $2\sim$ 8). For top-hat jet model, it also can fit observations (i.e., large $\theta_{obs}$ and the low observed luminosity) with a large $\theta_j$ (e.g., $\theta_{j}$$>$$10^{\circ}$), but a rather high local (i.e., z$<$0.01) short GRB rate ($>$8.1$\times10^3$) remains a problem for the top-hat model, when compared to the estimated rate (90$\sim$1850 $ Gpc^{-3}$$yr^{-1}$) in Fong et al. (2015) from a number of short GRBs. Another big challenge for top-hat jet model is the estimated $E_{peak,0}$($>$7.5$\times10^4$keV) of GRB 170817A, which is too large for a typical short GRB($\sim500$keV). The estimated $E_{peak,0}$ almost rules out the top-hat jet model. So We conclude that GRB 170817A is more likely an intrinsically low luminosity GRB ($10^{49}$ erg s$^{-1}$) with a structured jet. More observations can provide further information of the jet energy distribution of the similar low-luminosity short GRBs.

%Recently, Fermi GBM detected a short GRB~170817A after GW170817.  The host galaxy NGC~4993 of the event is 0.0098, resulting in the isotropic-equivalent \gr~energy, E$ _{iso,\gamma}$, of GRB~170817A of only about 4 $\times$ 10$^{46}$ erg.  

\section*{Acknowledgments}
We acknowledge the usage of Fermi GBM data, provided by the Fermi Science Support Center. XBH and PHT are supported by National Natural Science Foundation of China (NSFC) grants 11633007 and 11661161010, and RFS is supported by NSFC (11673078).
%In the uniform jet model with a more or less universal jet half-opening angle $\theta_j\approx0.1$ as considered by other authors (e.g., Jin et al. 2017). 

%%%%%%%%%%%%Begin the Reference%%%%%%%%%%%%%%%%%%%%%%%%%%\

   \clearpage
    \begin{figure*}
    \includegraphics[width=8.4cm, height=5.4cm, angle=0]{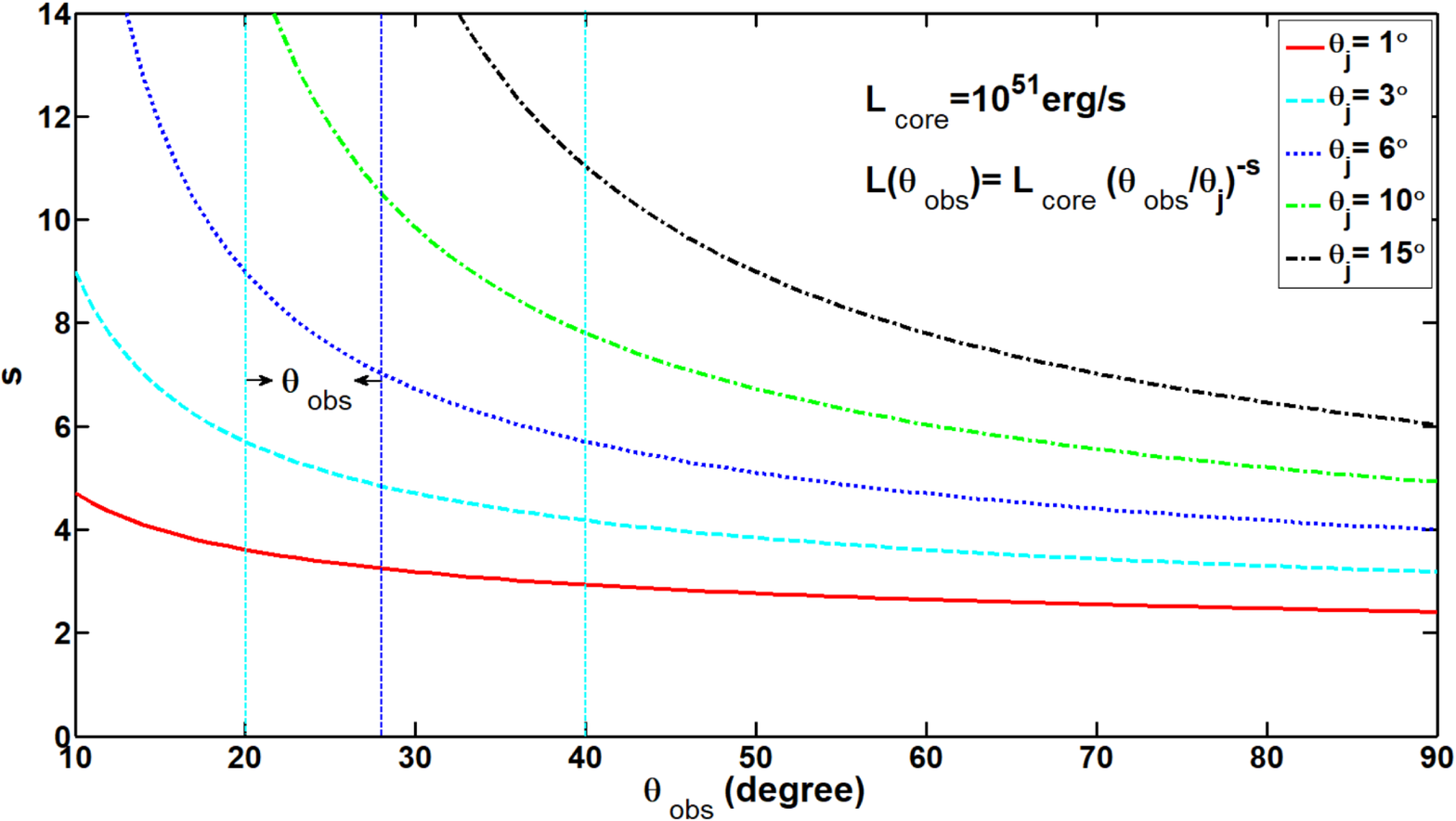}
    \includegraphics[width=8.4cm, height=5.4cm, angle=0]{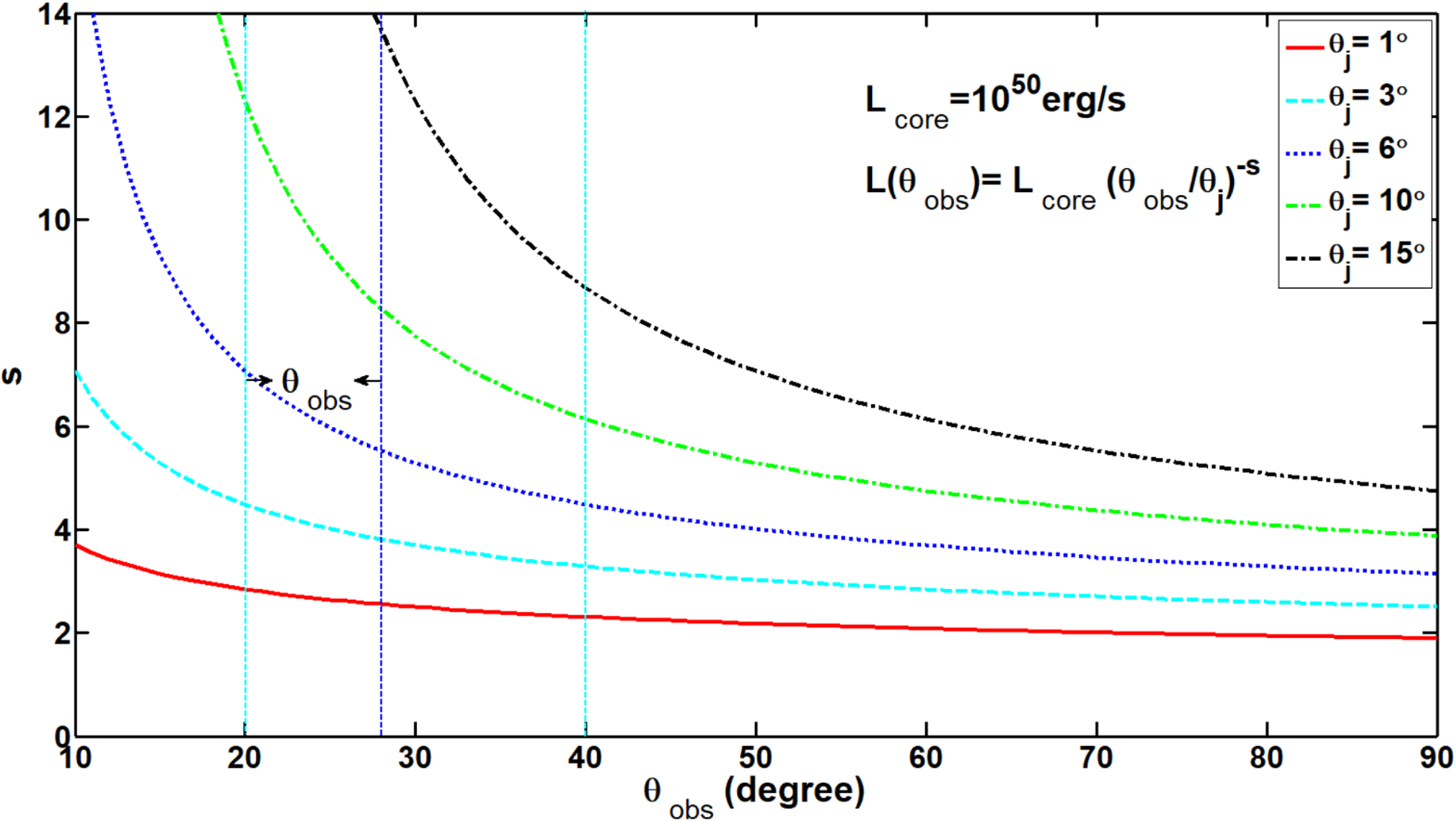}
    \end{figure*}   
    \begin{figure*}
    \centering
    \includegraphics[width=8.4cm, height=5.4cm, angle=0]{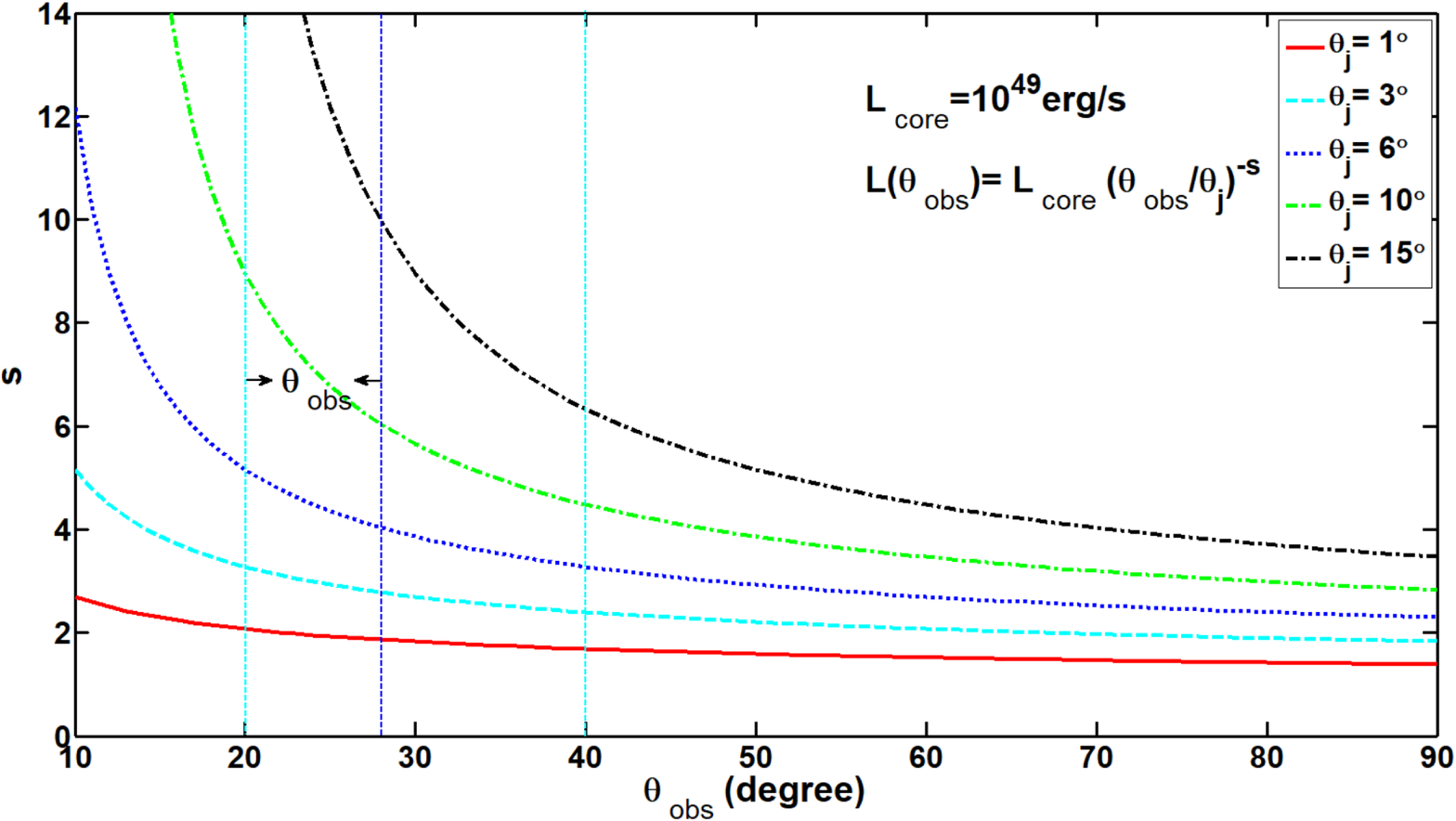}
      \caption{The constraint on $s$ and $\theta_{\rm obs}$ for an off-axis structured jet model for GRB 170817A. Here the angular profile of structured jet luminosity is $L(\theta_{\rm obs})= L_{\rm core} (\theta_{\rm obs}/\theta_j)^{-s}$ for off-axis observing angle $\theta_{\rm obs} > \theta_j$. We take the values $4\pi L(\theta_{\rm obs})= 2\times10^{46}$ erg s$^{-1}$ and the $4\pi L_{\rm core}$ is $10^{49}$ erg s$^{-1}$ to $10^{51}$ erg s$^{-1}$ for typical short GRBs. The dashed blue vertical lines are the viewing angles inferred from GW observation ($\theta_{\rm obs}\le28^\circ$; Abbott et al. 2017). The dashed cyan vertical lines are the viewing angle derived from afterglow modeling of radio and X-ray observation ($20^\circ\le\theta_{\rm obs}\le40^\circ$; Margutti \etal 2017). The  likely viewing angle is then plotted in the figures.}
         \label{theta_obs}
   \end{figure*}
\end{document}